\documentclass[%
reprint,
superscriptaddress,
amsmath,amssymb,longbibliography,
aps,prb
]{revtex4-2}
\usepackage[utf8]{inputenc}
\usepackage[T1]{fontenc} 
\usepackage{graphicx}
\usepackage{dcolumn}
\usepackage{bm}
\usepackage{amsmath}
\usepackage{lmodern}
\usepackage{mathtools}
\usepackage{float}
\usepackage[x11names]{xcolor}
\definecolor{PRX}{RGB}{46,48,146}
\definecolor{MyBlack}{RGB}{35,31,32}
\usepackage[colorlinks,citecolor=PRX  ,linkcolor=PRX,urlcolor=PRX]{hyperref}
\usepackage{siunitx}
\usepackage{chemformula}

\usepackage{chemformula} 
\usepackage[T1]{fontenc} 
\usepackage{soul}

\usepackage[normalem]{ulem}

\usepackage[normalem]{ulem}

\usepackage{titlesec}
\titleformat{\section}[display]{\bf\large}{}{0pt}{}
\titleformat{\subsection}[display]{\bf}{}{0pt}{}
\titlespacing\section{0pt}{12pt plus 4pt minus 2pt}{0pt plus 2pt minus 2pt}
\titlespacing\subsection{0pt}{12pt plus 4pt minus 2pt}{0pt plus 2pt minus 2pt}
\titlespacing\subsubsection{0pt}{12pt plus 4pt minus 2pt}{0pt plus 2pt minus 2pt}

\begin{document}

\title[Electrical switching of a chiral lasing from polariton condensate in a Rashba-Dresselhaus regime]
  {Electrical switching of a chiral lasing from polariton condensate in a Rashba-Dresselhaus regime}

\author{Karolina \L{}empicka-Mirek}
\affiliation{Institute of Experimental Physics, Faculty of Physics, University of Warsaw, ul.~Pasteura 5, PL-02-093 Warsaw, Poland}
\author{Mateusz Kr\'ol}
\affiliation{Institute of Experimental Physics, Faculty of Physics, University of Warsaw, ul.~Pasteura 5, PL-02-093 Warsaw, Poland}
\author{Luisa De Marco}
\affiliation{CNR NANOTEC, Institute of Nanotechnology, Via Monteroni, 73100 Lecce, Italy}
\author{Annalisa Coriolano}
\affiliation{CNR NANOTEC, Institute of Nanotechnology, Via Monteroni, 73100 Lecce, Italy}
\affiliation{Dipartimento di Matematica e Fisica E. De Giorgi, Università del Salento, Campus Ecotekne, Via
Monteroni, Lecce 73100, Italy}
\author{Laura Polimeno}
\affiliation{CNR NANOTEC, Institute of Nanotechnology, Via Monteroni, 73100 Lecce, Italy}
\author{Ilenia~Viola}
\affiliation{CNR-NANOTEC, Institute of Nanotechnology, UOS Rome, SLIM Lab
c/o Dip. Fisica, Università "La Sapienza", Piazzale A. Moro 2, 00185 - Rome, Italy}
\author{Mateusz K\k{e}dziora}
\affiliation{Institute of Experimental Physics, Faculty of Physics, University of Warsaw, ul.~Pasteura 5, PL-02-093 Warsaw, Poland}
\author{Marcin Muszy\'nski}
\affiliation{Institute of Experimental Physics, Faculty of Physics, University of Warsaw, ul.~Pasteura 5, PL-02-093 Warsaw, Poland}
\author{Przemys\l{}aw Morawiak}
\affiliation{Institute of Applied Physics, Military University of Technology, Warsaw, Poland}
\author{Rafa\l{} Mazur}
\affiliation{Institute of Applied Physics, Military University of Technology, Warsaw, Poland}
\author{Przemys\l{}aw~Kula}
\affiliation{Institute of Chemistry, Military University of Technology, Warsaw, Poland}
\author{Wiktor Piecek}
\affiliation{Institute of Applied Physics, Military University of Technology, Warsaw, Poland}
\author{Piotr Fita}
\affiliation{Institute of Experimental Physics, Faculty of Physics, University of Warsaw, ul.~Pasteura 5, PL-02-093 Warsaw, Poland}
\author{Daniele Sanvitto}
\affiliation{CNR NANOTEC, Institute of Nanotechnology, Via Monteroni, 73100 Lecce, Italy}
\author{Jacek Szczytko}
\author{Barbara Pi\k{e}tka}
\email{Barbara.Pietka@fuw.edu.pl}
\affiliation{Institute of Experimental Physics, Faculty of Physics, University of Warsaw, ul.~Pasteura 5, PL-02-093 Warsaw, Poland}

\begin{abstract}
Efficient optical classical and quantum information processing imposes on light novel requirements: chirality with low threshold non-linearities. In this work we demonstrate a chiral lasing from an optical modes due to emerging photonic Rashba-Dresselhaus spin-orbit coupling (SOC). For this purpose we developed a new electrically tunable device based on an optical cavity filled with birefringent liquid crystal (LC) and perovskite crystals. Our novel method for the growth of single crystals of CsPbBr\textsubscript{3} inorganic perovskite in polymer templates allows us to reach a strong light-matter coupling regime between perovskite excitons and cavity modes, and induce polariton condensation. The sensitivity of the LC to external electric fields lets us to tune the condensate energy in situ and induce synthetic SOC. This shapes the condensate between a single linearly polarized or two circularly polarized separated in momentum, emitting coherent light. The difference in the condensation thresholds between the two SOC regimes can be used to switch on and off the chiral condensate emission with a voltage.
\end{abstract}

\maketitle

\begin{figure*}
	\centering
	\includegraphics{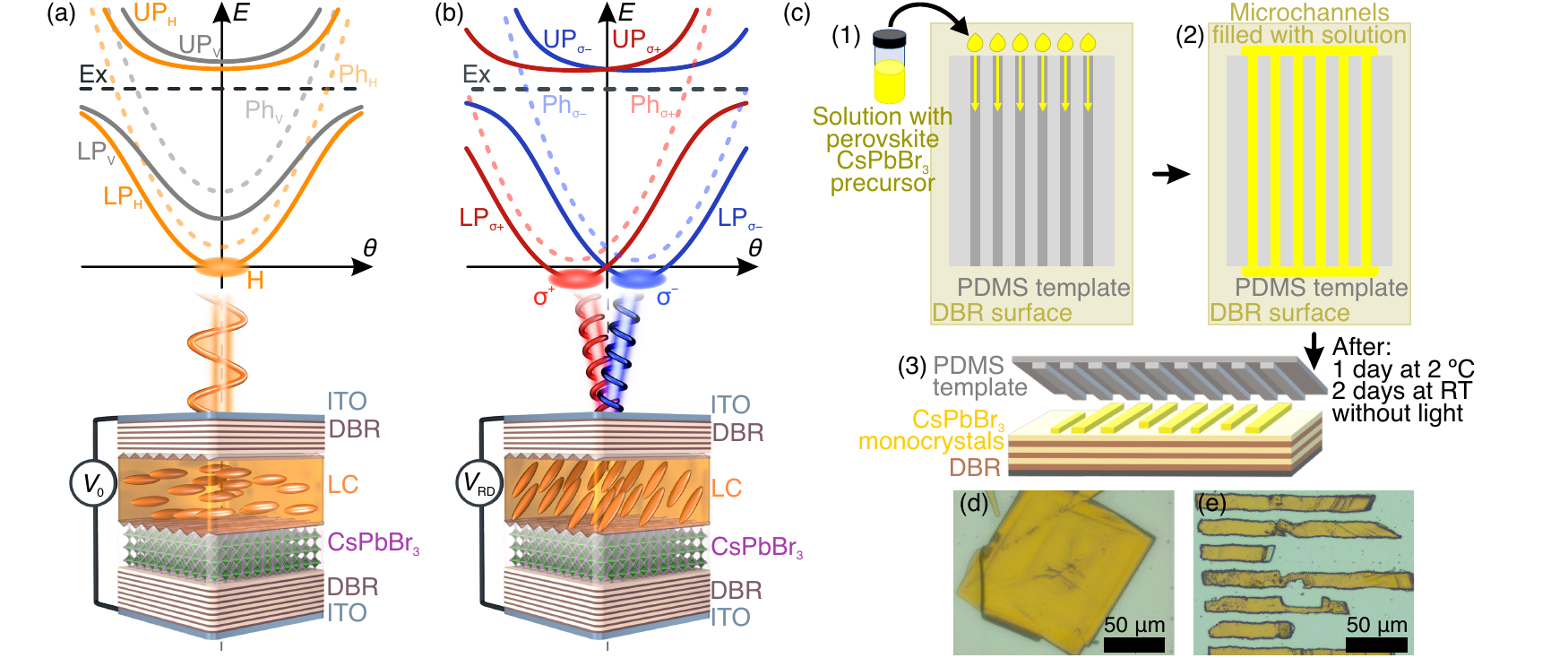}
	\caption{Liquid crystal microcavity polariton laser and synthesis of perovskite monocrystals. Scheme of the device and corresponding dispersion relation of polaritonic modes in two regimes. (a) Without external voltage $V_\text{0}$, anisotropy of the liquid crystal (LC) distinguishes H and V polarized modes, with the condensation occurring at the bottom of H mode. (b) With external voltage $V_\text{RD}$ corresponding to Rashba-Dresselhaus spin orbit coupling regime the dispersion relation consists of set of  two circularly polarized modes, separated in the emission angle. Condensation occurs simultaneously at the two degenerate minima of the dispersion relation. (c) Schematic of the synthesis of a perovskite by solution crystallization. (d--e) Optical microscope images of CsPbBr$_{3}$ perovskite monocrystals (d) 2~$\mu$m and (e) 500~nm thick.}
	\label{rys:idea}
\end{figure*}

The engineering of devices that allow for control over the polarization state of the emitted light is one of the fundamental objectives of nanophotonics. Specifically, spin-orbit interactions can be used to couple the polarization (spin) of a photon with its direction of propagation~\cite{Bliokh_NatPhoton2015}. Of special interest for rapidly developing fields such as light-matter interactions and quantum technologies are devices working with chiral light \cite{Lininger_AdvMater2022,Lodahl_Nature2017,Huebener_NatMater2020}. There are multiple different approaches to achieve circularly polarized emission, with special emphasis on lasing. The chirality of the emitted light can be imposed by the asymmetric design of the entire structure \cite{Yu_SciRep2020,Liao_SciRep2016,Maksimov_PRApp2022}. A planar cavity can be constructed from chiral mirrors for example based on metamaterials \cite{Plum_APL2022}, or cholesterics liquid crystals \cite{Mitov_AdvMater2012}. Lasing in structures using chiral organic molecules can exhibit a non-zero degree of circular polarization in the emitted light \cite{Yokoyama_AdvMater2006,Qu_ACSNano2021}. In so-called spin lasers circularly polarized emission is obtained from fully homogeneous structures by injection of spin-polarized carriers~\cite{Zutic_SolStatCommun2020}.  

Realistic applications of photonic devices in classical and quantum information processing strongly benefit from the nonlinear response of the device. One of the means of introducing strong interactions into an optical structure is to couple the photonic mode with an excitonic dipole. Emerging in the strong coupling regime mixed exciton polaritons states exhibit strong interactions; support polariton lasing or polariton condensation~\cite{Kavokin_NatRevPhys2022,Bloch_NatRevPhys2022}; superfluidity \cite{Amo_NatPhys2009}; long propagation distances and optical switching \cite{Ballarini_NatCommun2013}. In addition, planar microcavities show a wide range of spin-orbit coupling phenomena~\cite{Leyder2007,Gianfrate2020,Ren_NatCommun2021,Ren_LPR2022}.

Exciton polariton physics, initially investigated predominantly at low temperatures, thanks to progress in material science, is accessible at room temperature \cite{Ghosh_PhotIns2022}. A specific class of materials that have recently gained significant attention for broad optoelectronic applications are perovskites \cite{Fu_NatRevMater2019,Kim_ChemRev2020,Liu_NatMater2021}, in which the optical response exhibit pronounced excitonic effects \cite{BaranowskiPlochocka_AdvEnerMater2020}. Perovskites have been considered as a promising platform for room temperature polaritonics \cite{Su_NatMater2021}, due to unique phenomena observed in those materials e.g.: strong interactions \cite{Fieramosca2019}, polariton condensation \cite{Su_NatPhys2020,Spencer2021,Tao_NatMater2022}, parametric amplification~\cite{Wu_AdvPhot2021}, long distance propagation \cite{Su_SciAdv2018}, optical switching \cite{Feng_SciAdv2021} and H-V splitting \cite{Polimeno_Optica2021,Spencer2021,Su_SciAdv2021,Polimeno_NatNanotech2021,Tao_NatMater2022}.

In this work, we introduce inorganic \ch{CsPbBr3} perovskite monocrystals into an optical microcavity filled with a birefringent liquid crystal to demonstrate a strong coupling regime and emission from a polariton condensate. Thanks to the tunability of the LC optical anisotropy, we are able to control spin-polarized polariton modes with an external electric field \cite{LempickaMirek_SciAdv2022}. In addition, thanks to the birefringent LC medium, the cavity modes can be coupled to each other to obtain photonic Rashba-Dresselhaus spin-orbit coupling \cite{Rechcinska_Science2019}. This allows us to switch off linearly polarized light from the condensate or shift the emission into two circularly polarized beams, which exit the cavity at different angles \cite{Muszynski_PRApp2021,Li_NatCommun2022,Long_AdvSci2022}.

A scheme of the device is presented in Fig.~\ref{rys:idea}. The perovskite crystal is enclosed between two dielectric distributed Bragg reflectors (DBRs). Without the external electric field (Fig.~\ref{rys:idea}a) optically anisotropic LC molecules distinguish photonic cavity modes with horizontal (H) and vertical (V) polarization. In the strong coupling regime, the cavity modes mix with excitonic resonance in the perovskite, forming upper (UP\textsubscript{H/V}) and lower (LP\textsubscript{H/V}) polariton modes. Under pulsed excitation, the polaritons can form a condensate that occupies the bottom of the dispersion relation of the H-polarized mode.

Depending on the electric bias applied to the ITO electrodes, we can tilt the anisotropy axis of the LC layer. This changes the effective refractive index for horizontally polarized light, which affects the energy of H-polarized modes, where the V-polarized ones are stationary. In particular, such tunability can lead to a situation where two cavity modes with perpendicular polarizations are close to degenerate. In a specific situation where both modes are of opposite parity, the optical Rashba-Dresselhaus spin-orbit coupling between the modes emerges \cite{Rechcinska_Science2019}, with a characteristic splitting between circularly polarized modes (Fig.~\ref{rys:idea}b). The dispersion minima split into a degenerate pair at a nonzero emission angle. Above the condensation threshold, the emission from the cavity consists of two circularly polarized beams. Here, we additionally utilize the change of the condensation threshold between different regimes to switch on or off the condensate with the external field.

\section{Results and discussion}

\subsection{Preparation of a liquid crystal microcavity with perovskite single crystals}

Single crystals of inorganic CsPbBr$_{3}$ perovskite were synthesized directly on DBR substrates by means of a microfluidics-assisted technique on a confined microstructured polymeric template. This technique allows to regulate the solvent evaporation rates and hence the supersaturation levels of the perovskite solution confined within a microchannel template. In this way, it is possible to precisely finalize the nucleation process, control the crystal growth, and hence the dimension and quality of the obtained micro-crystals. A sketch of the synthetic process is shown in Fig.~\ref{rys:idea}c: a microchannel polymeric template of polydimethylsiloxane (PDMS) is placed in close contact with the DBR substrate and soaked with 3~$\mu$l of a 0.42~M solution of CsPbBr$_{3}$ in dimethyl sulfoxide (DMSO). Due to capillary action, the solution fills the microchannels. Initially, the crystallization process is observed under an optical microscope for the first 10 minutes after the infiltration of the solution. The surface of the dielectric mirrors is rough because of the polycrystallinity of the sputtered layers. Such a roughness, combined with the confinement of precursor solution inside the microchannels, favors the formation of nucleation centers of the perovskite, and from there the growth of crystals on the mirror surface begins.
After the appearance of the first perovskite crystals, the sample is sealed in a small box and left undisturbed at 2~$^{\circ}$C for 24~h and then at room temperature for 48~h in the dark under atmospheric conditions. At the end of this process, the PDMS template is removed from the substrate leaving perovskite single crystals on DBR, with the sizes of the patterned template. CsPbBr$_{3}$ crystals have a cubic structure; therefore, they prefer the self-organization of the crystals in the shape of thin cuboids. The obtained single crystals are of good quality, smooth, and homogeneous. One of the advantages of this technique is that it allows to carefully modulate the lateral dimensions (width and length) and thickness of the crystals by simply changing the PDMS template. For example, we have developed crystals with lateral dimensions of 50--60~$\mu$m with a thickness of 2~$\mu$m (Fig.~\ref{rys:idea}d) or crystals with lateral dimensions of 15~$\mu$m, a length of 20--70~$\mu$m and a thickness of 500~nm (Fig.~\ref{rys:idea}e).

Single perovskite crystals in the bottom dielectric mirror are protected with a 60~nm protective PMMA layer (3\% PMMA solution in anisole) applied using the spin-coating process. A polymer layer SE-130 is applied to the upper mirror (6 pairs of SiO$_{2}$/TiO$_{2}$ layers with ITO). Both mirrors with polymer layers are placed in order. The dielectric mirrors are assembled in a microcavity with a thickness of several $\mu$m, and a highly birefringent liquid crystal ($\Delta n = 0.4$) is forced into the cavity. Electrodes are soldered to the microcavity in order to rotate the molecules with the voltage, thereby achieving tuning of the polariton modes obtained in the cavity.

\subsection{Strong coupling regime}

	\begin{figure}
		\centering
		\includegraphics[width=\columnwidth]{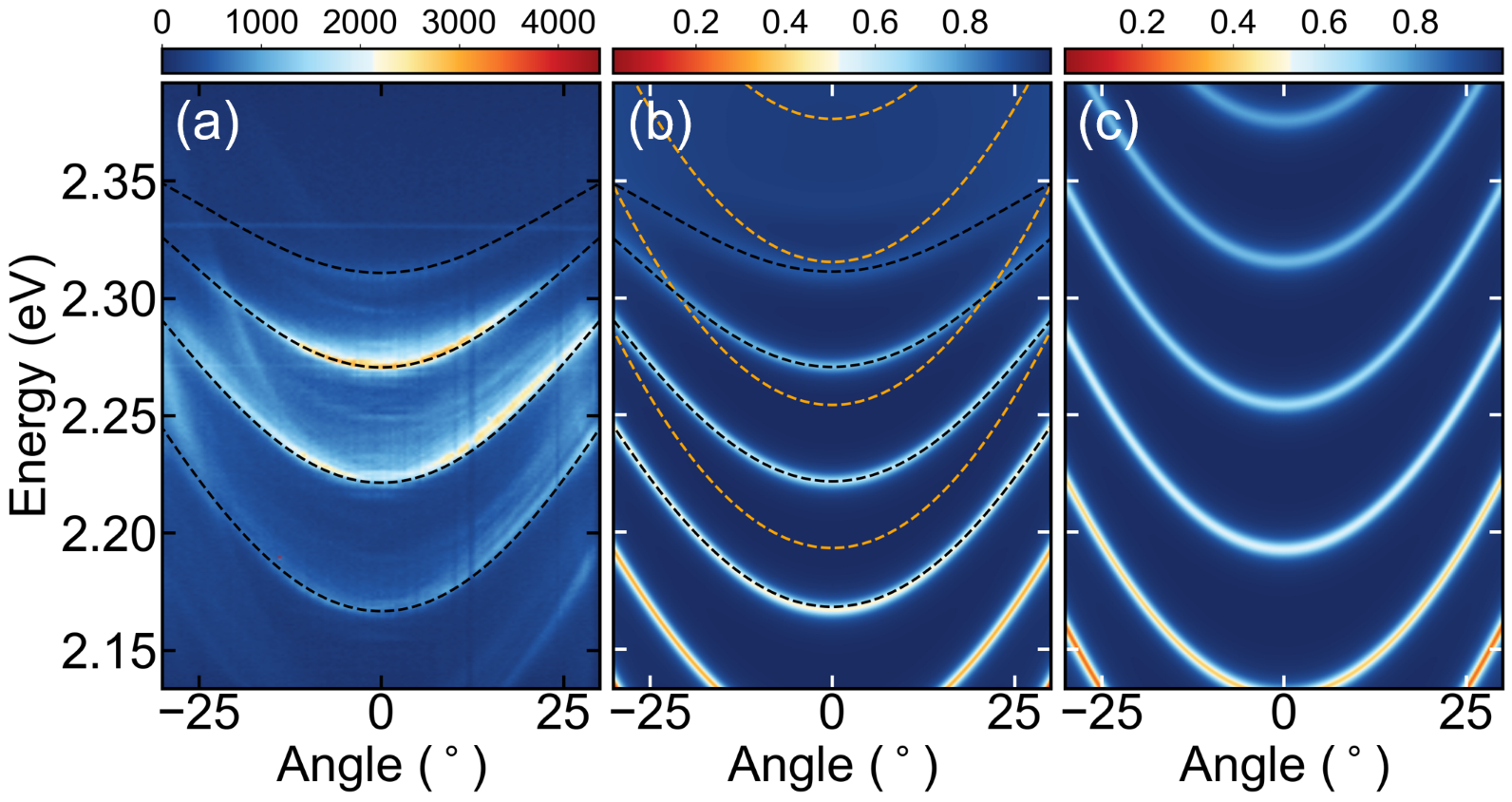}
		\caption{Strong coupling regime in microcavity with \ch{CsPbBr3} monocrystals. (a) Angle-resolved photoluminescence spectra from the microcavity. The black dashed lines mark the simulated dispersion of the structure presented in (b). (b) Simulated reflectance from the cavity with the perovskite. Orange dotted lines mark dispersion of the bare cavity modes and black dashed lines represent results from the coupled oscillator model. (c) Simulated reflectance from the corresponding structure with perovskite without excitonic resonance.}
		\label{PerLCMC_SC}
	\end{figure}

 	\begin{figure}
		\centering
		\includegraphics[width=\columnwidth]{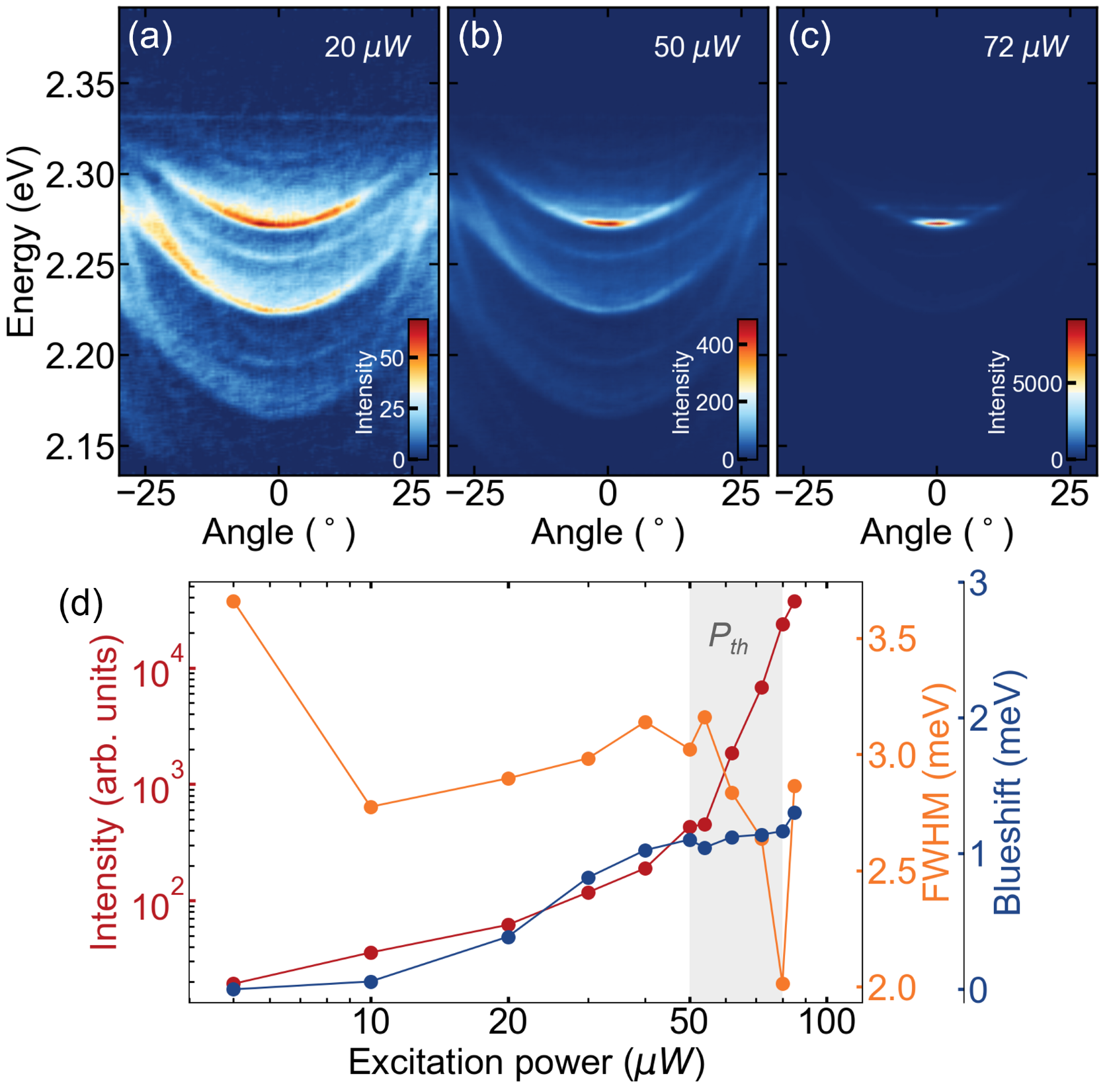}
		\caption{Polariton condensation from CsPbBr$_{3}$ perovskite LC microcavity.  (a--c) Angle-resolved photoluminescence spectra with polariton dispersion relation (a) below, (b) at, and (c) above threshold. (d) Polariton emission intensity, FWHM (full width at half maximum) and energy blueshift for increasing excitation power.}
		\label{rys:condensate_v3}
	\end{figure}

\begin{figure*}
	\centering
	\includegraphics[width=.99\textwidth]{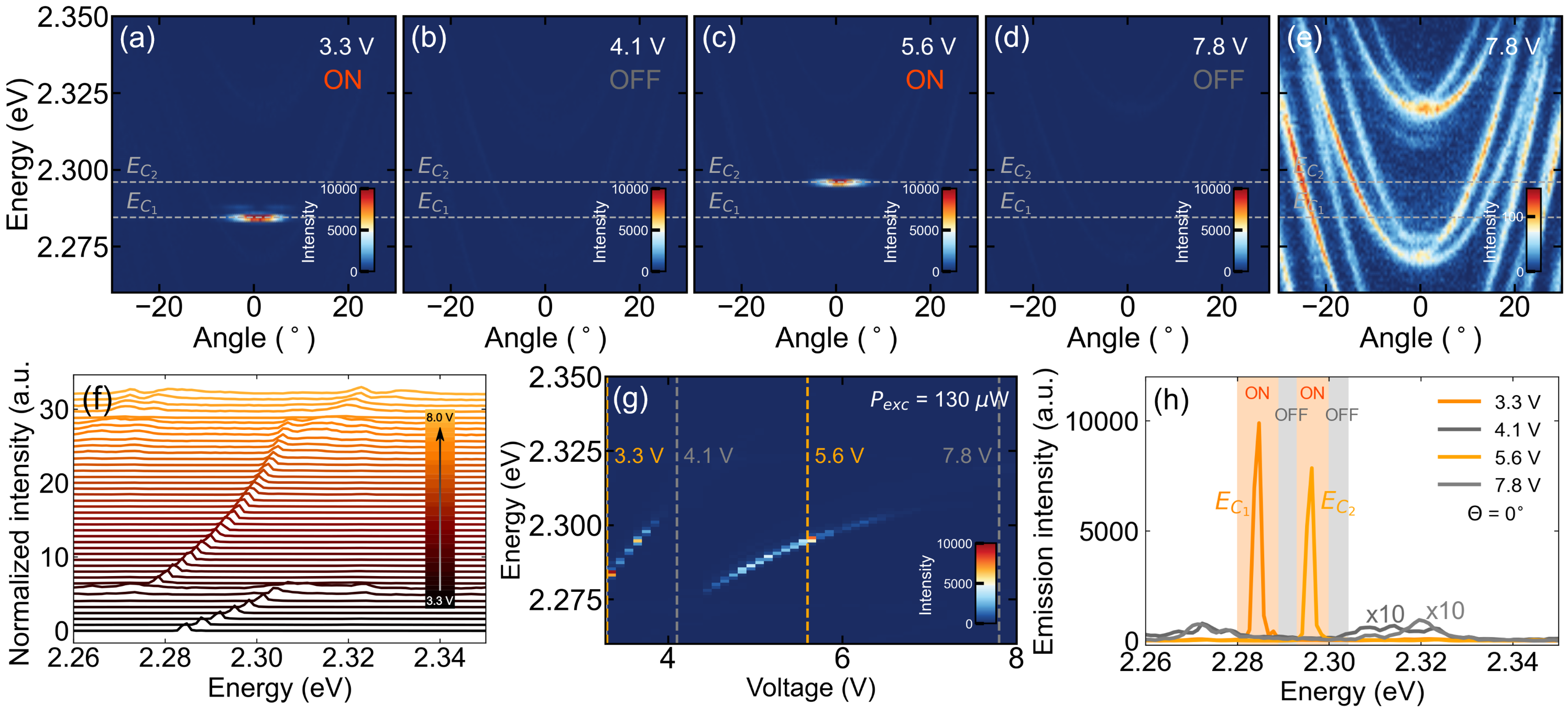}
	\caption {Electrical switching of a polariton condensation. (a--d) Angle-resolved photoluminescence spectra for different applied voltages with fixed intensity scale. Polariton condensation for 3.3~V (a) and 5.6~V (c). Weak emission spectra for 4.1~V (b) and 7.8~V (d). (e) The same as (d) but with rescaled intensity revealing Rashba-Dresselhaus SOC regime. (f,h) Emission intensity at a normal angle for increasing voltage illustrating smooth energy tunning of polariton condensate. (g) Normal angle emission spectra for (a--d), showing the switching of polariton condensation ON/OFF.}
\label{rys:switching_BEC_v1}
\end{figure*}

To demonstrate a strong coupling regime between the perovskite exciton and cavity modes, we investigated areas containing thick monocrystals of CsPbBr\textsubscript{3}. Fig.~\ref{PerLCMC_SC}a presents angle resolved photoluminescence spectra collected at excitation powers (with a pulsed laser) significantly below the condensation threshold. Because of the high thickness of the perovskite monocrystal, the emission spectra consist of multiple modes. The strong coupling regime can be observed by the curvature of the observed modes. The lowest energy mode (2.167\,eV at normal incidence) exhibits a parabolic dispersion relation with respect to the observation angle. This is an expected behavior for the predominantly photonic mode, in which the energy detuning from the exciton is significantly higher than the coupling strength. For higher energy modes, the effective mass visibly increases. For the highest energy mode visible in the emission (2.311\,eV) the dispersion relation shows a significantly higher effective mass originating from the anticrossing behavior with the excitonic transition (2.4\,eV). As a result of high absorption of the perovskite at energies close and above the exciton line, the upper polariton branches are not observed in the emission.

The experimentally observed dispersion of the modes is in perfect agreement with the numerical simulations. The optical response of a layered, optically anisotropic structure can be calculated using the Berreman method. Fig.~\ref{PerLCMC_SC}b presents simulated reflectance from 4.95\,$\mu$m thick perovskite enclosed between two \ch{SiO2}/\ch{TiO2} DBRs using this approach. To simulate the perovskite layer, we used the dielectric function extracted from the ellipsometric data from Ref.\,\cite{Zhao_JMChC2018}. 
The deeps in the simulated reflectance correspond to the polaritonic modes in the structure. Their position was fitted and plotted on top of the photoluminescence results (Fig.~\ref{PerLCMC_SC}a) by black dashed line. The simulated dispersion precisely follows the experimental data. 

To show that the observed dispersion of the modes can be considered as a result of the strong coupling between the exciton in the perovskite crystal and the cavity mode, additional numerical modeling was performed. The parameters of the dielectric function of the perovskite allow the removal of the peak corresponding to the excitonic resonance. Simulations performed for such structure lacking the excitonic resonance are presented in Fig.~\ref{PerLCMC_SC}c. As expected, the spectrum consists of a series of parabolic bare cavity modes. Their dispersion relation can be used in a coupled oscillator model to mix with the exciton state to obtain the exciton-polariton modes. The lower polariton branches resulting from four independent 2$\times$2 oscillator models with exciton resonance at 2.4\,eV and consecutive bare cavity modes are presented in Fig.~\ref{PerLCMC_SC}b in black dashed lines. All calculated for the same coupling strength of $\Omega_\text{R} = 76.4$\,meV accurately follow the simulated deeps in the reflectance (Fig.~\ref{PerLCMC_SC}b), as well as the experimental photoluminescence results (Fig.~\ref{PerLCMC_SC}a).

\subsection{Polariton condensation in the LC microcavity}

The presence of a strong coupling regime between the perovskite excitons and the cavity modes allows the investigation of nonlinear phenomena in the system. Fig.~\ref{rys:condensate_v3} demonstrates polariton condensation in a perovskite liquid crystal cavity, without applying an electric bias to the electrodes. At low powers of non-resonant pulsed excitation (2.95 eV, 150~fs pulse duration) the emission follows the polaritonic modes, as presented on angle-resolved photoluminesce spectra in Fig.~\ref{rys:condensate_v3}a. With increasing pump power, the emission concentrates at the bottom of the dispersion relation (Fig.~\ref{rys:condensate_v3}b--c), which corresponds to the emission at a normal angle to the cavity plane. The observed dependence of the emission intensity, spectral line width, and emission energy with pumping power---presented in Fig.~\ref{rys:condensate_v3}d---follows the typical behavior for the formation of a polariton condensate. Above the condensation threshold of 72~$\mu$W the emission intensity nonlinearly increases,  followed by an abrupt decrease of the emission linewidth. The steady increase of the emission energy with the excitation power can be interpreted as a result of polariton-polariton interactions stemming from the excitonic component of the polariton.

\subsection{Electrical switching of a polariton condensate}

Our optically anisotropic cavity confines photonic modes, corresponding to the directions parallel (H-polarized) and perpendicular (V-polarized) to the LC director. Condensation occurs preferentially in the H-polarized mode \cite{Muszynski_PRApp2021}, the energy of which can be tuned by an external voltage applied to the ITO electrodes in the sample. For a selected position in the sample, we determined the condensation threshold at 130~$\mu$W without the external electric bias. Fig.~\ref{rys:switching_BEC_v1}a presents an angle-resolved photoluminescence spectrum just above the condensation threshold power, where most of the emission is concentrated close to 0 deg angle. Keeping the excitation power constant, with increasing voltage up to 4.0~V, by shifting the polaritonic modes, we can increase the energy of the condensate from 2.278~eV to 2.309~eV (energy tuning of 31~meV in total), as shown in Fig.~\ref{rys:switching_BEC_v1}f,g presenting the normal incidence emission spectra. With a further increase of the voltage, the polariton mode approaches in energy another one with perpendicular V-polarization. Due to the emerging optical spin orbit effect coupling the two modes, the condensation threshold goes beyond the excitation power, and the laser-like emission is turned off, as shown in Fig.~\ref{rys:switching_BEC_v1}b. Above 4.2~V another H-polarized mode approaches the optimum gain spectral range, and the condensate switches on (Fig.~\ref{rys:switching_BEC_v1}c), and can be further energy-tuned up to 7~V. Once again above that value, the H-polarized mode approaches the V-polarized one, resulting in low emission intensity (Fig.~\ref{rys:switching_BEC_v1}d). Photonic Rashba-Dresselhaus coupled modes are clearly visible in Fig.~\ref{rys:switching_BEC_v1}e (rescaled Fig.~\ref{rys:switching_BEC_v1}d). Here, for coupled modes, the condensation threshold is higher than the excitation power, and the emission comes from all states along the dispersion relation. The general behavior of the condensate is summarized in Fig.~\ref{rys:switching_BEC_v1}f--h, presenting emission spectra at a normal angle for different voltages. Depending on just the amplitude of the external electric field, we are able to energy-tune the emission from the condensate or to switch it off. 

\subsection{Circularly polarized polariton condensation}

 \begin{figure}
	\centering
	\includegraphics[width=\columnwidth]{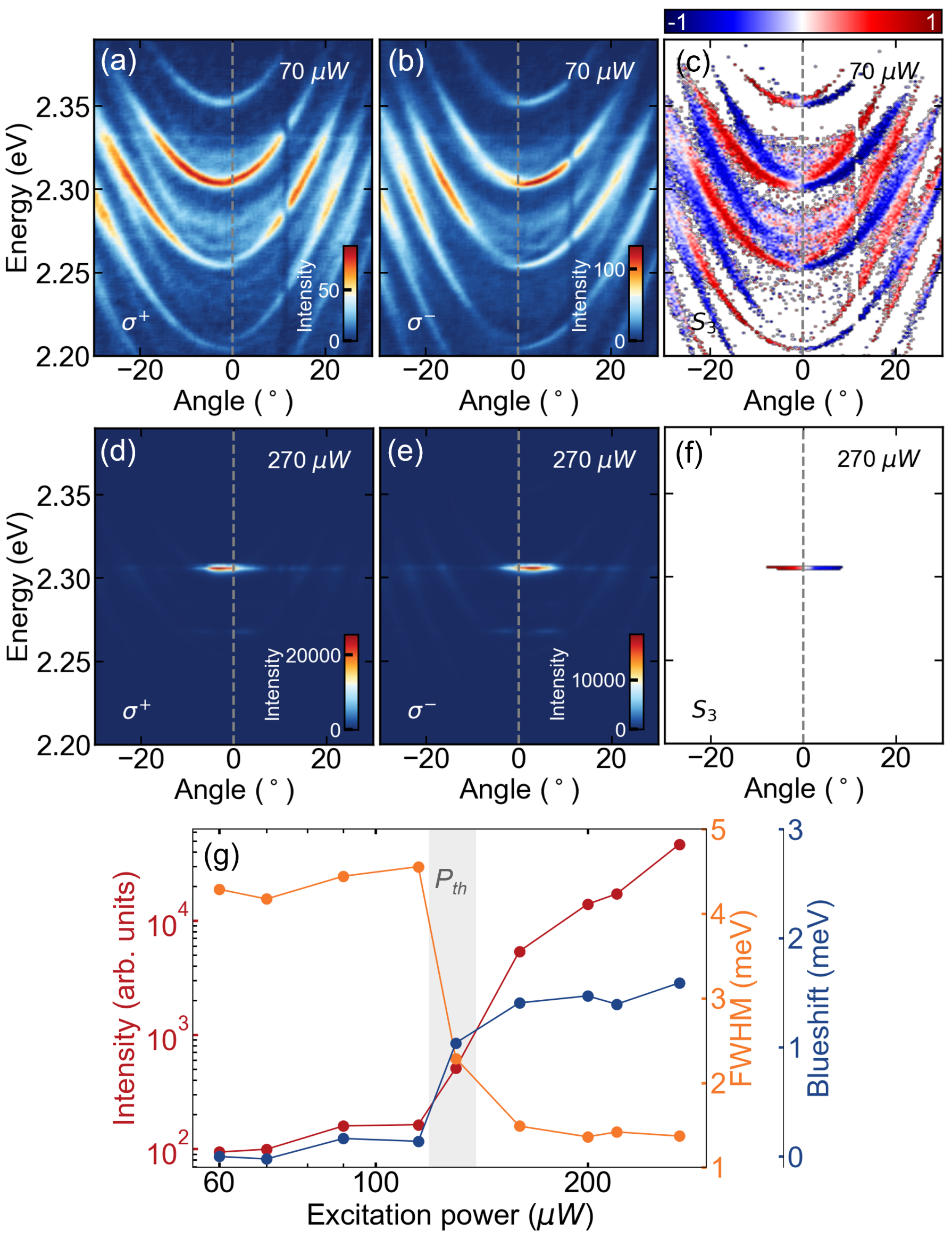}
	\caption{Polariton condensation in the Rashba-Dresselhaus regime. Angle-resolved photoluminescence spectra in both circular polarizations and resulting the $S_3$ parameter: (a--c) below and (d--f) above the condensation threshold. (g)~Emission intensity, spectral linewidth and energy blueshift for increasing excitation power.}
	\label{rys:RD_condensate}
\end{figure}

Polariton condensation in the Rashba-Dresselhaus spin-orbit coupling regime can be observed in Fig.~\ref{rys:RD_condensate}. To reach this regime, an external voltage of 5.3~V is applied to the device. Angle-resolved photoluminescence maps resolved in circular polarizations below the condensation threshold in Fig.~\ref{rys:RD_condensate}a clearly demonstrate two degenerate modes. The full circular polarization of the modes and their separation in the emission angle can be observed by plotting the third Stokes parameter [$S_3 = \left({I_{\sigma^+}-I_{\sigma^-}}\right)/\left({I_{\sigma^+}+I_{\sigma^-}}\right)$, where $I_{\sigma^+}$ ($I_{\sigma^-}$) is the intensity of $\sigma^+$ ($\sigma^-$) polarized light].

With increased excitation power to 130~$\mu$W, two circularly-polarized condensates form at a nonzero emission angle, corresponding to the energy minima of the dispersion relation, as presented in Fig.~\ref{rys:RD_condensate}d-f. The observed dependence of the emission intensity, spectral linewidth and the energy of the condensate on the excitation power, shown in Fig.~\ref{rys:RD_condensate}g, follows the typical behavior of a polariton condensate. The observed threshold power equal to 160~$\mu$W is approximately two times higher than in the situation without the external electric field.  

Polariton condensation is typically regarded as a non-equilibrium process, limited by kinetics. When occupation at some state exceeds 1, due to Bosonic stimulation the condensates build up. When the linearly polarized modes of the cavity are energy-separated, that process occurs at the bottom of the H-polarized mode (Fig.~\ref{rys:idea}a). When the cavity modes are coupled with the optical Rashba-Dresselhaus term, the dispersion relation exhibits two minima that correspond to the opposite handedness of the emitted light, as shown in Fig.~\ref{rys:idea}b. Both of those minima are degenerate, so they are equally populated with increasing excitation power. To reach the Bosonic stimulation regime, occupation of both of those states must approach unity at the same time, which effectively increases the condensation threshold when compared to a situation in which a single mode can drain most of the energy provided by the excitation beam.

\section{Conclusions}

To summarize, we proposed a new method to obtain single crystals of inorganic CsPbBr\textsubscript{3} perovskite. The growth process that takes place within a polymer template, under ambient conditions, can be used for the preparation of high-quality crystals with controlled size. The perovskite prepared in this way, was placed in an optical microcavity filled with a nematic liquid crystal. The high optical quality of the material, in conjunction with a strong coupling regime between excitons in the perovskite and photonic cavity modes, allowed for the creation of polariton condensates at room temperature. In our device, the laser-like emission from the condensate can be spectrally tuned by 31~meV, thanks to the sensitivity of the liquid crystal to an external voltage. In addition, we used birefringence of the liquid crystal to induce the Rashba-Dresselhaus spin-orbit coupling.  It resulted in chiral polaritonic modes and above the threshold in simultaneous emission from the polariton condensate in two, tilted beams exhibiting polarization of opposite handedness. As the different regimes of spin-orbit interactions differ in the condensation threshold, we showed that it can be additionally operated as an electrically controlled switch of the emission.

\section{Methods}

\subsection{Synthesis of monocrystals of inorganic perovskite CsPbBr$_{3}$ } 

Single crystals were obtained by slow microfluidics-assisted crystallization from a perovskite precursor solution. A 0.42~M solution was prepared from the mixture of reagents: 106 mg of CsBr and 183.5~mg of PbBr$_{2}$ (Molar ratio 1:1) were added to 1.2~ml of DMSO. This solution was mixed at 80~$^{\circ}$C on the hot plate in a glove box, in a nitrogen atmosphere for 2 hours. 
The resulting yellow-clear solution was taken from the glove box, where perovskite single crystals were subsequently crystallized under atmospheric conditions. The crystals were crystallized in microwire-shaped polydimethylsiloxane (PDMS, Sylgard 184, Dow-Corning) templates made by a conventional soft lithography process. In this case, two types of microchannels were used: a single wide rectangular channel 150~$\mu$m wide and 2~$\mu$m high and a network of microchannels 15~$\mu$m wide and 500~nm high. A volume of 3~$\mu$l of yellow perovskite precursor solution is deposited at the inlet of the PDMS template, patterned both with single microchannel and a network, and 
placed in conformal contact with a 8 DBR SiO$_{2}$/TiO$_{2}$ dielectric mirror (SiO$_{2}$ on top) with ITO (indium tin oxide). After the appearance of fine perovskite crystals, the sample is kept in a box, opportunely closed with parafilm, and then transferred to a refrigerator for 24~h and stored at a constant temperature of 2~$^{\circ}$C. The presented crystals crystallized after 3 days, during which two consecutive days the process was carried out under atmospheric conditions and room temperature but without the sample's access to light.

\subsection{Photoluminescence measurements}

The photoluminescence spectra were excited non-resonantly at 2.95\,eV (420\,nm) with
sub-150\,fs pulses generated by an amplified Ti-sapphire system working at
the repetition rate of 5\,kHz. The incident laser beam was focused on the sample using a 50$\times$ microscope objective with $\text{NA} = 0.65$. The emitted light was collected with the same objective. Angular resolution of the emission signal was obtained by imaging of the Fourier plane of the collecting objective onto an entrance slit of a spectrometer equipped with a CCD sensor. 

\subsection{Voltage applied to the cavity}

Steering of the LC was performed with a square waveform with frequency of 1\,kHz and variable amplitude from a function generator.

\section{Acknowledgments}

This work was supported by the National Science Centre grants 2019/35/B/ST3/04147, 2018/31/N/ST3/03046 and 2020/37/B/ST3/01657; NAWA Canaletto grant PPN/BIT/2021 /1/00124/U/00001 and the European Union’s Horizon 2020 program, through a FET Open research and innovation action under the grant agreement No. 964770 (TopoLight).

\bibliography{bibliography}

 \end{document}